\documentclass[12pt]{spieman}  
\usepackage{amsmath,amsfonts,amssymb}
\usepackage{graphicx}
\usepackage{setspace}
\usepackage{tocloft}
\usepackage{lineno}

\title{Improved noise performance from the next-generation buried-channel p-Mosfet SiSeROs}

\author[a]{Tanmoy Chattopadhyay}
\author[a]{Sven Herrmann}
\author[a]{Matthew Kaplan}
\author[a]{Peter Orel}
\author[b]{Kevan Donlon}
\author[c]{Gregory Prigozhin}
\author[a]{Glenn Morris}
\author[b]{Michael Cooper}
\author[c]{Andrew Malonis}
\author[a,d,e]{Steven W. Allen}
\author[c]{Marshall W. Bautz}
\author[b]{Chris Leitz}

\affil[a]{Kavli Institute of Astrophysics and Cosmology, Stanford University, 452 Lomita Mall, Stanford, CA 94305, USA}
\affil[b]{MIT Lincoln Laboratory, Lexington, MA, USA}
\affil[c]{Kavli Institute for Astrophysics and Space Research, Massachusetts Institute of Technology, Cambridge, MA, USA}
\affil[d]{SLAC National Accelerator Laboratory, 2575 Sand Hill Road, Menlo Park, CA 94025, USA}
\affil[e]{Department of Physics, Stanford University, 382 Via Pueblo Mall, Stanford CA 94305, USA}

\cftpagenumbersoff{figure}
\cftpagenumbersoff{table}

\begin{document} 
\maketitle

\begin{abstract}
 The Single electron Sensitive Read Out (SiSeRO) is a novel on-chip charge detector output stage for charge-coupled device (CCD) image sensors. Developed at MIT Lincoln Laboratory, this technology uses a p-MOSFET transistor with a depleted internal gate beneath the transistor channel. The transistor source-drain current is modulated by the transfer of charge into the internal gate. At Stanford, we have developed a readout module based on the drain current of the on-chip transistor to characterize the device. In our earlier work, we characterized a number of first prototype SiSeROs with the MOSFET transistor channels at the surface layer. An equivalent noise charge (ENC) of around 15 electrons root mean square (RMS) was obtained. In this work, we examine the first buried-channel SiSeRO. We have achieved substantially improved noise performance of around 4.5 e$^-_{RMS}$ and a full width half maximum (FWHM) energy resolution of 132 eV at 5.9 keV, for a readout speed of 625 kpixel/s. We also discuss how digital filtering techniques can be used to further improve the SiSeRO noise performance. Additional measurements and device simulations will be essential to further mature the SiSeRO technology. This new device class presents an exciting new technology for the next-generation astronomical X-ray telescopes requiring fast, low-noise, radiation-hard megapixel imagers with moderate spectroscopic resolution.
\end{abstract}

\keywords{SiSeRO, X-ray detector, readout electronics, digital filtering, instrumentation}

{\noindent \footnotesize\textbf{*}Tanmoy Chattopadhyay,  \linkable{tanmoyc@stanford.edu} }

\begin{spacing}{1}   


\section{INTRODUCTION}
\label{sec:intro}  
CCDs \cite{Lesser15_ccd,gruner02_ccd,ccd_janesick01} 
have been the primary detector technology for soft X-ray instrumentation for more than three decades in X-ray astronomy. The low noise performance ($<$4 $e^{-}_{RMS}$), broad energy response of up to fifteen keV, high spatial and moderate energy resolutions made them an obvious choice for the X-ray spectro-imagers in missions such as the Advanced Satellite for Cosmology and Astrophysics (\emph{ASCA}), \emph{Chandra}, the X-ray Multi-Mirror Mission (\emph{XMM-Newton}), the Neil Gehrels Swift Observatory, \emph{Suzaku}, \emph{Hitomi}, and the \emph{AstroSat} mission. \emph{Chandra}, in particular, has been extremely successful in revealing the detailed properties of X-ray astronomical sources by combining high angular resolution imaging with good spectral resolution and the excellent quantum efficiency  of CCDs. 

Next-generation X-ray CCDs, with higher readout speeds and improved noise performance, also hold the potential to support future astronomical missions such as the \emph{Lynx} Observatory \cite{gaskin15_lynx}, proposed earlier to NASA's 2020 Decadal Survey, and the Advanced X-ray Imaging Satellite (\emph{AXIS}) probe class mission\footnote{https://axis.astro.umd.edu/}, which will have order-of-magnitude larger collecting areas to probe deeper into the high redshift, faint X-ray universe. Noise performance reaching in the sub-electron regime (i.e. below 1 $e^{-}_{RMS}$) and the associated improvements in the low energy response ($<$500 eV) will also be highly beneficial to fully utilize these observatories. MIT Lincoln Laboratory (MIT-LL), MIT and Stanford University (SU) have together made substantial improvements in X-ray CCD technology in recent years \cite{bautz18,bautz19,bautz20}, and in the development of fast, low noise readout electronics to support these detectors \cite{chattopadhyay22_ccd,herrmann20_mcrc,Bautzetal2022,Oreletal2022}. In parallel,
MIT-LL has been developing a unique Single electron Sensitive Read Out stage (hereinafter SiSeRO \cite{chattopadhyay22_sisero,Chattopadhyayetal2022}) for CCDs, intended to provide an even greater responsivity (pA/$e^{-}$) and significantly better noise performance at readout speeds $>$2 MHz. 

In Chattopadhyay et al. 2022 \cite{chattopadhyay22_sisero}, the working principle of SiSeROs was discussed and the first results from these devices were demonstrated using a readout module developed at SU. The working principle of SiSeROs is similar in some respects to DEPFET detectors \cite{kemmer87_depfet,strueder00_depfet_imager} and draws from earlier work on floating-gate amplifiers described in Matsunaga et al. 1991 \cite{matsunaga91}.  
For our first generation prototypes, we obtained an equivalent noise charge (ENC) of 15 $e^{-}_{RMS}$. In this paper, we 
present results for a new SiSeRO device (CCID-93) with a buried-channel p-MOSFET transistor, obtaining highly encouraging results. We have also developed techniques to suppress 1/f correlated noise using optimized digital filters when measuring the charge signal in the pixels.    

In Sec. \ref{sec:sisero}, we give a brief introduction to the SiSeRO amplifiers. This is followed in Sec. \ref{sec:results} by a short description of the characterization test stand, readout module and the new test results for a CCID-93 buried-channel SiSeRO using standard analysis approaches. In Sec. \ref{sec:digfiltering}, we present and discuss improvements obtained with the application of new digital filtering techniques. The results are summarized along with future plans in Sec. \ref{summary}.


\section{Overview of SiSeRO devices}\label{sec:sisero}

\begin{figure}
    \centering
    \includegraphics[scale=0.5,bb=200 0 400 400]{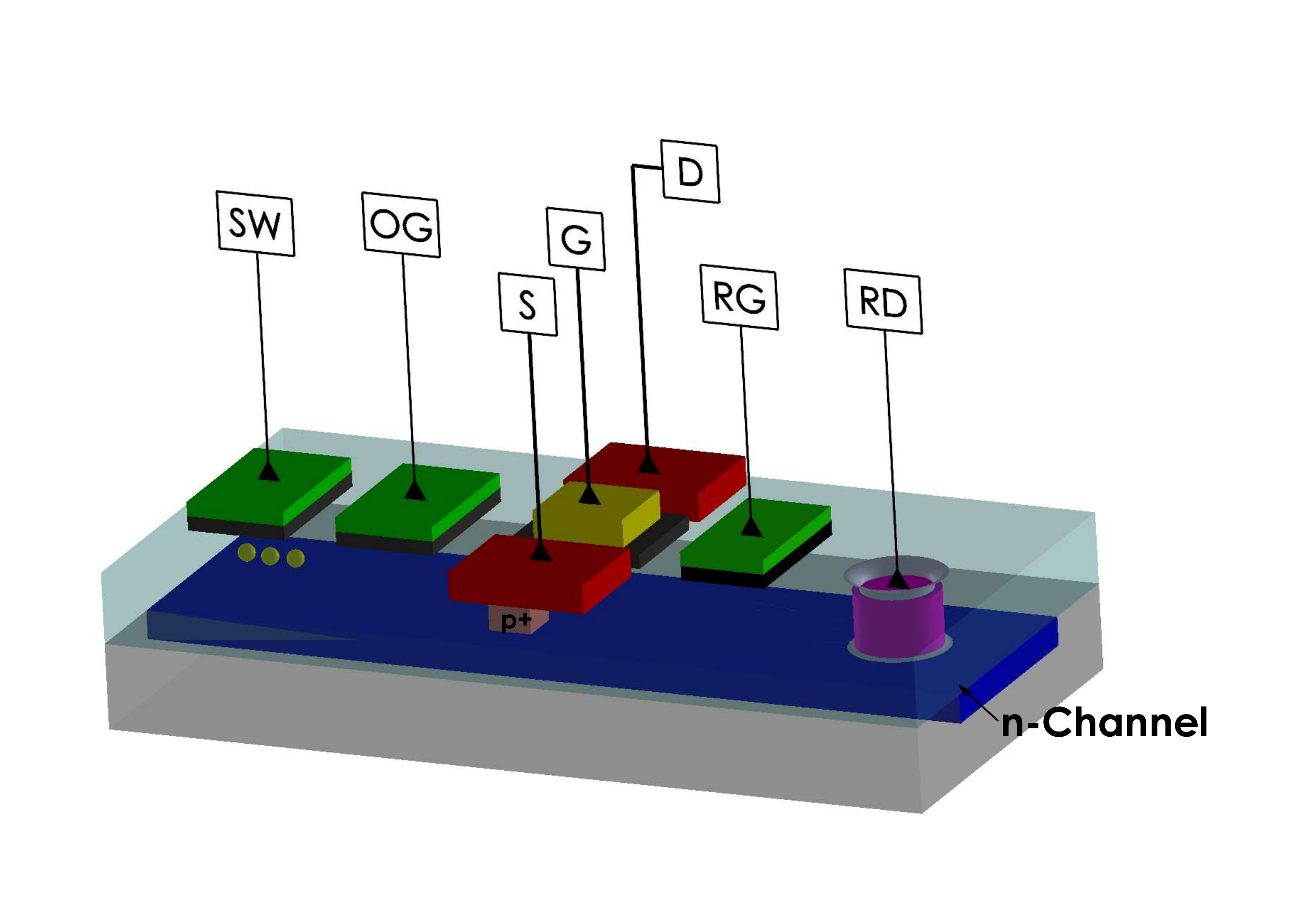}
    \caption{Overview 3D sketch of the buried-channel SiSeRO device. It uses a p-MOSFET transistor with an internal gate beneath the p-MOSFET. Charge transferred from the last serial gate (SW) of the serial register to the internal gate through the output gate (OG) modulates the drain current of the p-MOSFET. 
    }
    \label{sisero}
\end{figure}
Details of the SiSeRO working principle can be found in Chattopadhyay et al. 2022 \cite{chattopadhyay22_sisero}. The SiSeRO output stage, shown in Fig. \ref{sisero}, has a floating-gate amplifier, where a p-MOSFET gate (`S', `G', and `D' in the figure refer to the Source, Gate, and Drain respectively) sits above the CCD channel. When a CCD charge packet is transferred beneath the gate, it modulates the drain current of the transistor, in a manner proportional to the source signal strength. The signal is brought directly out of the detector through the drain of the transistor, enabling the device to be operated in a drain-current readout mode. Once the charge is read out, the internal channel is reset by emptying the channel to the reset drain (`RD') using the reset gate (`RG') switch. The  summing well (`SW') and `OG', in the figure refer to the last serial clock register and output gate respectively. The output gate is connected to a small positive DC bias to shift the charge from the serial register to the internal channel.   

For current readout devices like the SiSeROs, the thermal noise of the output stage depends on the conversion gain, $G_q$ (pA/electron) and the transconductance, $g_m$ ($\mu$S) of the p-MOSFET transistor\footnote{\url{https://www.nikhef.nl/~jds/vlsi/noise/sansen.pdf}}. Assuming only thermal noise for MOSFET transistors, it can be shown that the device output stage noise can be as low as 1 $e^{-}_{RMS}$ for a $G_q$ of 1400 pA/electron and $g_m$ of around 20 $\mu$S for 1 MHz of readout speed. In the case of the SiSeROs, the internal gate minimizes parasitic capacitance on the sense node, resulting in 
a high conversion gain and minimized noise. 
Further, MOSFETs provide larger transconductance per unit area than JFETs (used in CCDs) and especially the use of small MOSFETs can lead to very high signal speeds.  
These properties distinguish SiSeROs as a device class with the potential for very low noise and high-speed performance.

In Chattopadhyay et al. 2022 \cite{chattopadhyay22_sisero}, we used the first generation SiSeRO prototype devices to demonstrate their working principle and presented the first results using a drain current readout module. We measured a charge conversion gain of 700 pA/electron and around 20 $\mu$S of transconductance \cite{chattopadhyay22_sisero}, obtaining an ENC of 15 $e^{-}_{RMS}$ and a FWHM spectral resolution of 230 eV at 5.9 keV. These first prototype devices used surface-channel transistors in which we expect the silicon interface states to capture and release mobile carriers, causing excess noise at a level multiple times that of the theoretical lower limit. In this paper, we characterize a new SiSeRO device that uses a buried transistor channel. In the absence of trapping and de-trapping of charge carriers from the silicon interface states, we can expect improved noise performance.


\section{Characterization of a buried-channel SiSeRO device (CCID-93)}\label{sec:results}

\begin{figure}
    \centering
    \includegraphics[width=1\linewidth]{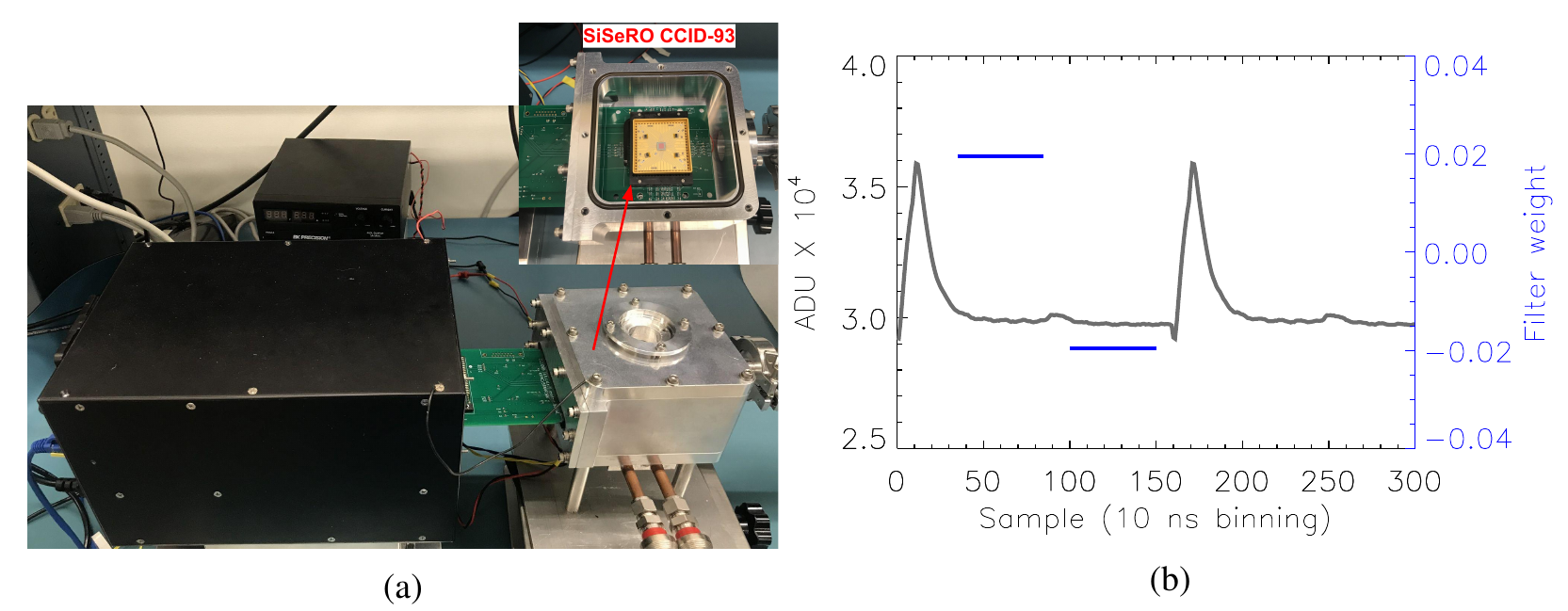}
    \caption{(a) The experiment set up with a CCID-93 bured channel SiSeRO device mounted inside the ``Tiny Box" vacuum chamber. A beryllium window mounted on the top flange (not shown here) is used for X-ray entrance window. (b) SiSeRO output video waveform (shown for 2 pixels here) obtained from the Archon controller. The blue lines represent the weight factors over the baseline and signal samples for the correlated double sampling (CDS). The difference between the two levels is proportional to the source signal.}
    \label{setup}
\end{figure}
\subsection{Experiment Test Stand}
The test stand, along with the readout module (also known as the `Tiny Box', see Chattopadhyay et al. 2020 \cite{chattopadhyay20_spie} for details) and the X-ray detector are shown in Fig. \ref{setup}a. A compact (13 cm $\times$ 15 cm $\times$ 6.5 cm) aluminum vacuum chamber houses the X-ray detector. The detector is mounted on an aluminum block. A thermo-electric cooler (TEC) is used to cool down the aluminum block and the detector. 
We use a Proportional-Integral-Derivative (PID) controller to control the temperature of the detector with better than 0.1$^\circ$C accuracy. An X-ray entrance window on the top flange allows X-ray photons to illuminate the detectors while maintaining the vacuum.

For this experiment, we used a CCID-93 detector (see Fig. \ref{setup}a), a prototype X-ray CCD with a buried-channel SiSeRO at the output stage. The test device, developed by MIT-LL, is fabricated in an n-channel, low-voltage, single-poly process. The device active area is $\sim$4 mm $\times$ 4 mm in size with a 512 $\times$ 512 array of 8 $\mu$m pixels. The test detector, used in the experiment, is a front-illuminated device.
The readout module consists of a preamplifier board (the green circuit board in Fig. \ref{setup}a) and an Archon controller \cite{archon14} (the black box in the figure). The preamplifier uses a drain current readout circuit where a current-to-voltage (I2V) amplifier converts and amplifies the SiSeRO output current to a voltage signal that is fed into a fully differential output driver. This driver converts the I2V single-ended output voltage to a fully differential signal for the differential analog to digital converter (ADC) in the Archon controller. In Chattopadhyay et al. 2022 \cite{chattopadhyay22_sisero}, we discuss the schematic of the preamplifier chain along with LTspice simulations\footnote{\url{https://www.analog.com/en/design-center/design-tools-and-calculators/ltspice-simulator.html}} for the transient and ac performances of the circuit. 
The Archon \cite{archon14}, procured from Semiconductor Technology Associates, Inc (STA\footnote{\url{http://www.sta-inc.net/archon/}}), provides the required bias and clock signals to run the detector, digitize the output signal, and perform CDS to generate 2D images. 

\subsection{Test results using standard analysis method (box filtering)}

The biasing condition of the SiSeRO is achieved by optimizing the reset clock (V$_\mathrm{RGLow}$ and V$_\mathrm{RGHigh}$), reset drain (V$_\mathrm{RD}$), output gate (V$_\mathrm{OG}$), and the p-MOSFET biases (V$_\mathrm{Source}$, V$_\mathrm{Drain}$, and V$_\mathrm{Gate}$). This is done in a two step process.
\begin{enumerate}
    \item First, we investigate the I-V characteristics of the SiSeRO for different values of V$_\mathrm{RGLow}$ and V$_\mathrm{OG}$. The source-drain current should be zero (I$_\mathrm{SD}=0$), when the V$_\mathrm{Gate}$ is below the threshold voltage for the transistor channel to form. This ensures that there is no parasitic path between the transistor source and drain through the surrounding gates (reset and output gate).
    \item V$_\mathrm{RGHigh}$ and V$_\mathrm{RD}$ are optimized to ensure complete reset of the internal channel, meaning  that there is complete charge removal and no back injection of charge into the internal channel after the reset. To test this, we perform one additional fast reset immediately after the charge transfer.
    This additional reset is supposed to drain the X-ray charge packets and for a perfect reset the baeline and the (now reseted) signal level are identical. We tune values of V$_\mathrm{RD}$ and V$_\mathrm{RGHigh}$ to ensure a proper reset of the channel such that there is no signal seen in the detector image.  
\end{enumerate}
Optimization of these parameters have been discussed in our previous work on SiSeROs \cite{chattopadhyay22_sisero} and the same approach has been followed here. For the tests here, we apply +5V, +1V, and +2V to the source, drain, and gate of the p-MOSFET respectively. At these biasing conditions, the transistor current, I$_\mathrm{SD}$, is around 50 $\mu$A.  

Figure \ref{setup}b shows the digital video waveform sampled at 10 ns intervals as obtained from the Archon controller. The readout speed of the detector is set at 625 kpixel/s (1600 ns to readout each pixel).
The change from the baseline (denoted by the horizontal extent of the first blue solid line from the left) to the signal level (denoted by the horizontal extent of the second blue solid line), after the charge packet is transferred from the output gate to the internal channel (see Fig. \ref{sisero}), is directly proportional to the amount of charge transferred.
The Archon controller extracts the signal amplitude by taking the difference between the sums of these two levels (digital correlated double sampling or DCDS) for each pixel, such that charge in j$^{th}$ pixel is given by 
\begin{equation}
    C_j^{CDS} = \sum\limits_{t_{0j}}^{t_{1j}}{h(t)~B_j(t)} ~+~ \sum\limits_{t_{2j}}^{t_{3j}}{-h^\prime(t)~S_j(t)},
    \label{eq:boxcds}
\end{equation}
where, $B_j$ and $S_j$ are j$^{th}$ baseline and signal samples respectively. $t_0$, $t_1$, $t_2$, $t_3$ define the limits of the time windows. $h(t)$ is the sampling or filter function used to sample the baseline and signal. $h^\prime(t)$ is the reverse of $h(t)$. In the standard method (by Archon controller), the baseline and signal levels are measured by using a constant sampling function, $h(t) = h^\prime(t) = Constant$. This is shown in the Fig. \ref{setup}b, where the blue solid lines represent the weight factors in sampling. This is known as differential average (similar to the dual-slope integrator method in the case of analog CDS \cite{hegyi80_dsi,ccd_janesick01}). Because of the constant multiplicative factor used for filtering, we hereafter refer to this method as digital box filtering.     

\begin{figure}
    \centering   
        \includegraphics[width=1\linewidth]{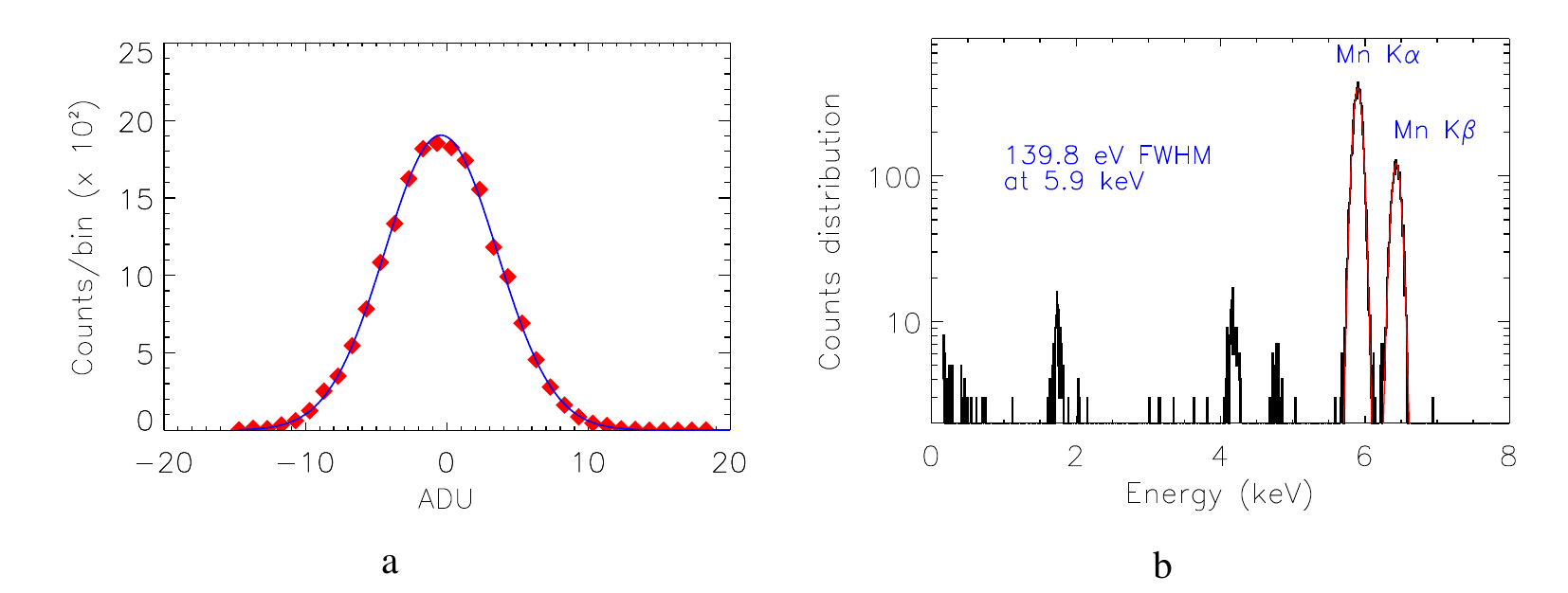} 
    \caption{Results using digital Box filtering: (a) distribution of charge (in ADU) from the over-scan region. The blue line is a Gaussian fit to the
distribution. The read noise (1$\sigma$ width of the Gaussian) is measured to be around 6.1 $e^{-}_{RMS}$. (b) Spectrum showing the Mn K$_\alpha$ (5.9 keV) and K$_\beta$ (6.4 keV) lines from a $^{55}$Fe radioactive source for
single-pixel (grade 0) events. The FWHM is measured to be 140 eV at 5.9 keV.}
    \label{results}
\end{figure}

Read noise is calculated from the distribution of signal in the overclocked pixels (an array of over-scanned pixels at the end of each pixel row), assuming that amount of signal charge there is negligibly small and the width of the histogram is entirely due to readout circuitry. 
An example of such distribution, obtained from one of the dark frames at 250 K (-23$^\circ$C), is shown in Fig. \ref{results}a.
The distribution is fitted with a Gaussian (shown in blue solid line) to quantify the RMS of the distribution. From the known conversion gain of the system, the read noise is measured to be around 6.1 $e^{-}_{RMS}$.  
Buried-channel SiSeROs, therefore, show a significant improvement in the noise performance compared to the earlier surface channel SiSeROs, where we measured the noise to be around 15 $e^{-}_{RMS}$ \cite{chattopadhyay22_sisero}. 

The spectral performance of the device was evaluated using a $^{55}\mathrm{Fe}$ radioisotope. The X-ray images are first corrected for bias (using the over-scan region) and dark current (using the dark frames). We generate an event list, where each event includes signal amplitudes of 9-pixels around every local maximum in spatial distribution (presumably due to X-ray interaction with silicon). The selection of such events is based on a primary threshold (7 times the read noise). We then apply a secondary threshold (2.6 times the read noise) to determine whether an event contains signal charge in the pixels surrounding the center and grade the events depending on number of pixels containing such extra charge. Spectra for each of the event grades are generated by adding charge in the adjacent pixels exceeding the secondary threshold.
An X-ray spectrum showing Mn K$_\alpha$ (5.9 keV) and Mn K$_\beta$ (6.4 keV) lines from the radioactive source is shown in Fig. \ref{results}b. The spectrum is generated using single-pixel events (grade 0 events), where the single-pixel events are those where none of the surrounding eight pixel charge amplitudes crosses the secondary threshold of 2.6 times the read noise. Therefore, the single-pixel events are considered to be generated from X-ray events depositing their full energy in one pixel. The red lines are Gaussian fits to the Mn K$_\alpha$ (5.9 keV) and Mn K$_\beta$ (6.4 keV) lines.
We calculate a FWHM of 140 eV at 5.9 keV. This energy resolution, in the case of buried-channel SiSeROs, represents a significant improvement over the previous results of 230 eV at 5.9 keV for the surface channel SiSeROs. We also calculated the conversion gain ($G_q$) of the system to be around 800 pA per electron. The transconductance ($g_m$) of the device is found to be around 25 $\mu$S. To measure transconductance, we calculate the ratio of drain current to gate voltage of the SiSeRO p-MOSFET transistor, $\mathrm{\Delta I_{Drain}~/~\Delta V_{Gate}}$. Drain current is measured from the detector circuit board for known gate voltages.


\section{Noise characterization of SiSeROs}\label{sec:digfiltering}

\subsection{Noise spectrum}
\begin{figure}
    \centering
    \includegraphics[width=0.7\linewidth]{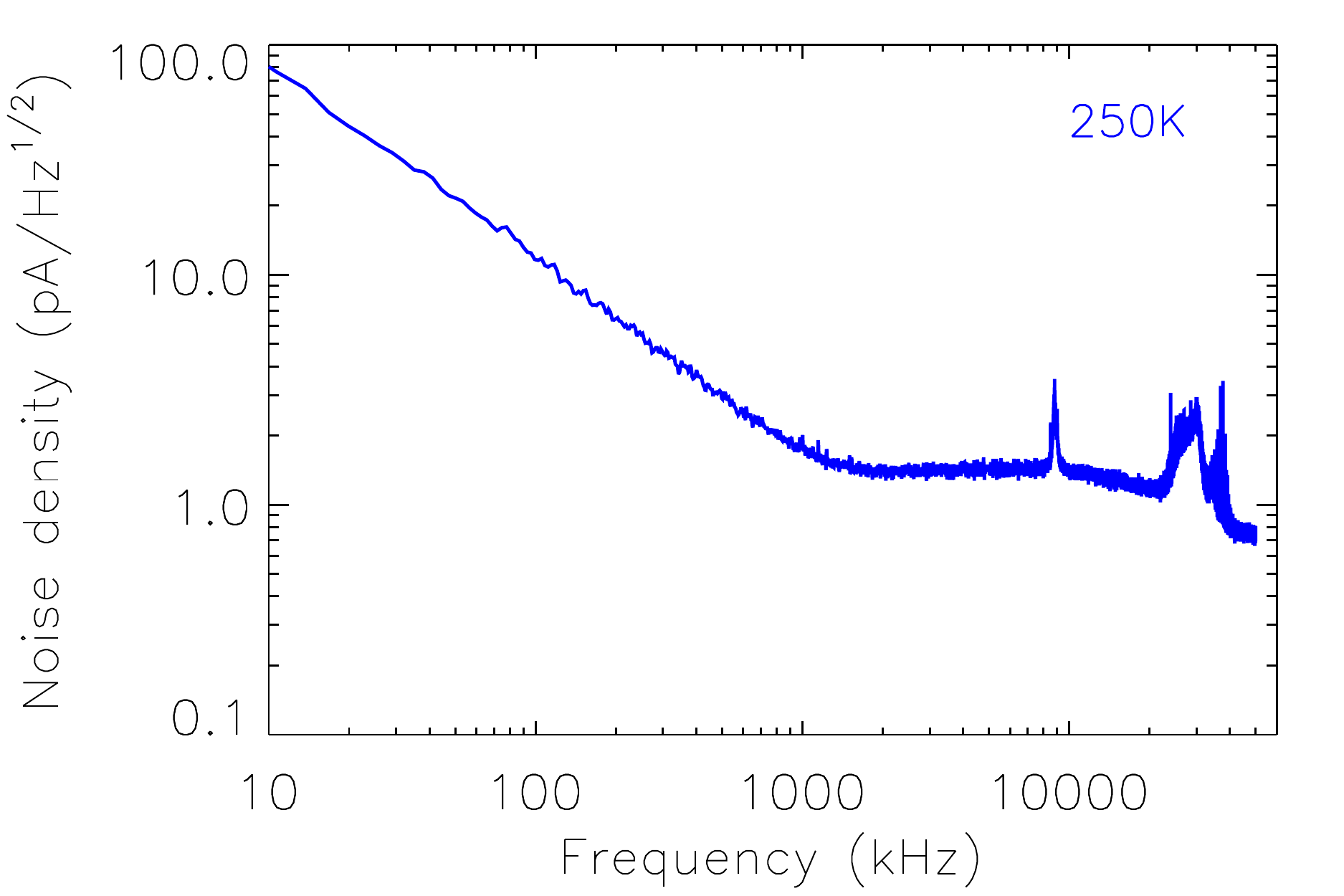}
    \caption{Input-referred noise density (in picoampere / $\sqrt{Hz}$) of the SiSeRO at 625 kHz readout speed and 250 K detector temperature. 
    The white thermal noise (flat part of the spectrum $>$1 MHz) floor is around 1.4 pA/$\sqrt{Hz}$. There is a pronounced 1/f noise at lower frequencies ($<$1 MHz). The short peaks seen at higher frequencies ($> 10 MHz$) are due to electromagnetic interference (EMI) from the turbo pump. EMI at those high frequencies are not expected to affect the noise because the effective gain of the readout chain is low at those frequencies.}
    \label{fig:noise_spectrum}
\end{figure}

\begin{figure}
    \centering
    \includegraphics[width=0.8\linewidth]{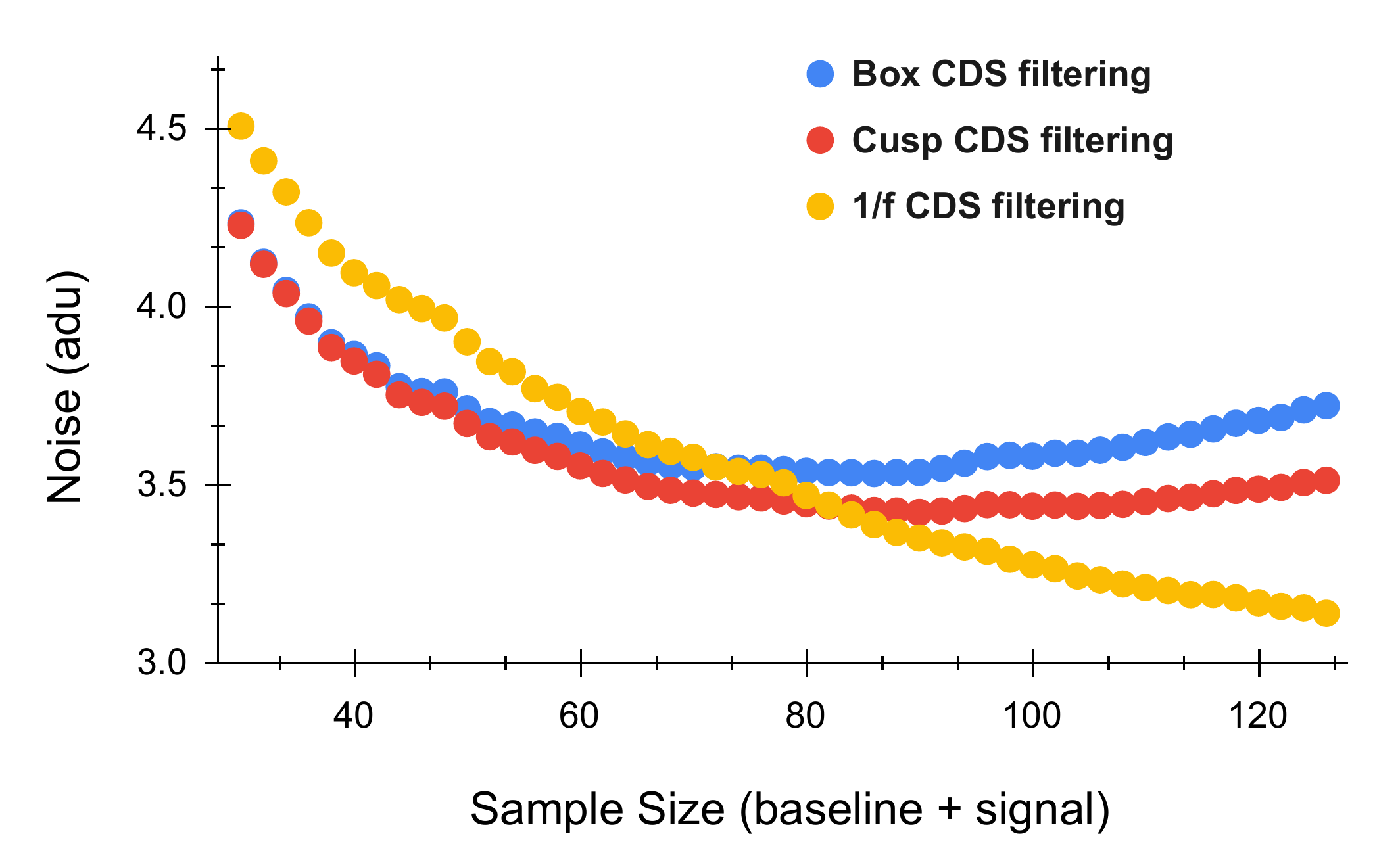}
    \caption{A comparison between the box CDS filtering (blue) and the optimized digital filtering methods (cusp in red and 1/f noise filter in yellow). At lower sample size or at the high frequency region of the noise spectrum (Fig. \ref{fig:noise_spectrum}), the total noise is dominated by the thermal noise. Therefore, with the increase in sample size, the noise decreases. When the sample size is large (low frequency region of the power spectrum), noise starts to increase again because of the dominance of 1/f noise. The 1/f digital filter outperforms the cusp and box filters because of the better 1/f noise reduction.}
    \label{fig:box_cusp}
\end{figure}

From the measured conversion gain of 800 pA per electron and transconductance of 25 $\mu$S, the theoretical limit in the thermal noise of these SiSeRO devices (with the readout electronics combined) is expected to be much lower than the measured 6.1 $e^{-}_{RMS}$ at a readout speed of 625 kpixel/s. In order to investigate the noise sources in the SiSeRO, we derived the noise power spectral density of the system from the time series ADC samples with no signal charge reaching the output. 
In Fig. \ref{fig:noise_spectrum}, we see a prominent 1/f noise slope (also known as correlated noise as it correlates among many pixels) in the frequency spectrum with 1/f corner frequency close or exceeding the readout rate of the detectors (625 kpixel/s in this case). 
The frequency response of a box filter can be obtained from FFT of Eq. \ref{eq:boxcds}, which is $\sim (1/f) \sin{f}$ (also known as a $Sinc$ function). 
The central or peak frequency of the filter (where the filter gain is maximum) is $f=1/T_{int}$, where $T_{int}$, the filter integration time, is the time span between the baseline and signal midpoint, e.g. 
\begin{equation}
   T_{int} = (t_2 - t_1) + (t_1-t_0)/2 + (t_3-t_2)/2, 
\end{equation}
which is around 1.5 MHz in this case and the width of the filter varies roughly as 1/N$_{sample}$. 
From the white thermal noise floor of 1.4 pA/$\sqrt{Hz}$ and using simple approximations in the filter bandwidth calculations, we expect a noise yield of 4-5 $e^{-}_{RMS}$ for a hypothetical scenario where the noise spectrum is dominated primarily by thermal white noise. 
However, in the presence of large 1/f noise as in this case (see Fig. \ref{fig:noise_spectrum}), a box filter is not able to achieve the minimum noise yield, even though its band-pass filter approaches zero in f$^{-1}$ trend at low frequencies. 
This is shown in Fig. \ref{fig:box_cusp}, where we demonstrate the noise yield for the digital box filtering method (and other digital filters to be discussed later) as a function of sample size. 
While a larger sample size or lower bandwidth of the filter is useful to minimize the contribution of thermal noise, the filter peak frequency also shifts to a lower frequency, thus, deeper into the 1/f domain. As a result, the noise (blue data points for box) starts to increase after a certain sample size because of the prevalence of the 1/f noise. This puts a limit on the useful sample size in the box CDS and therefore the noise performance of the detectors, although maximizing the sample size is preferred in order to minimize the thermal noise as much as possible.

In the case of CCDs, there exists various digital filtering techniques for CDS that have been shown to successfully suppress 1/f correlated noise \cite{Stefanov2015,Cancelo2012,gach03_1/f}. 
We utilize some of these techniques in an attempt to improve the noise performance of the detectors. In our test setup, we have the flexibility to save raw ADC samples from the Archon controller. We utilized the raw samples to explore the digital filtering techniques.

\subsection{Optimized digital filters for correlated double sampling}
\subsubsection{Cusp filtering}
First, we discuss the effect of an optimized filter (hereafter `cusp' \cite{Stefanov2015}) on the noise performance. The sampling function is given by 
\begin{equation}
    h(t) = 1 + \frac{P-1}{1+t~C},
\end{equation}
where, $P$ is the peak of the sampling function and $C$ is a parameter. Note, $P=1$ returns the box filtering function.
\begin{figure}
    \centering
     \includegraphics[width=1\linewidth]{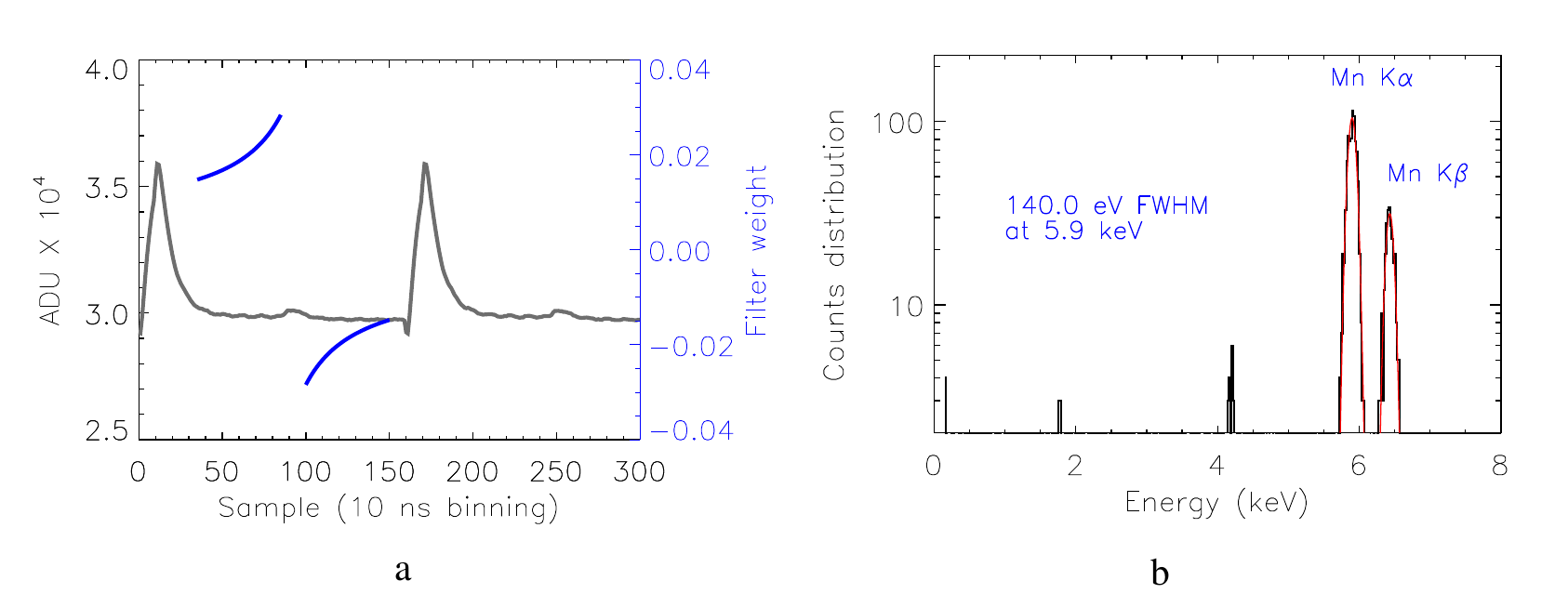}
    \caption{Results of ``cusp" digital filtering: (a) the blue lines (plotted against the right axis) show the weights in cusp filter used over the baseline and signal samples of the waveform. With this filtering technique, the read noise is improved to 5.8 $e^{-}_{RMS}$. (b) A single-pixel spectrum showing the Mn K$_\alpha$ (5.9 keV) and K$_\beta$ (6.4 keV) lines obtained using this method. The FWHM is measured to be 140 eV at 5.9 keV.}
    \label{fig:filter_cusp}
\end{figure}
This can be visualized in Fig. \ref{fig:filter_cusp}a, where the blue solid lines are the weighting factors on the baseline and signal samples respectively.
Because of the correlated 1/f noise, the baseline and signal levels within a pixel are affected by small but unequal amounts, which leads to additional noise after the CDS. To avoid this, the baseline and signal should be measured close in time. This type of filtering function applies more weight to samples close in time, and therefore reduces the noise contribution of signals with short term correlation. In the frequency domain, it can be seen that because of the larger weights near the transition, the central frequency of the filter is higher (therefore further away from the 1/f corner frequency) than the box filter for the same sample size, which helps in suppressing the 1/f noise more effectively than the box technique. This is shown in Fig. \ref{fig:box_cusp}, where we clearly see that with the cusp technique (red), we get better noise yield compared to box (blue).   

The final noise performance depends on multiple factors e.g. the sampling region (number of baseline and signal samples) and the values of $P$ and $C$, optimization of which in turn depends on the 1/f corner frequency (see Stefanov 2015 \cite{Stefanov2015} for more details). We optimize the parameters, $P$ and $C$, by studying the performance of the sampling function for different
parameters values and their influence on the output noise.
In our analysis, we fixed the values of $P$ and $C$ at 5.0 and 0.03 respectively. The read noise was improved slightly, by around 5 \%, to 5.8 $e^{-}_{RMS}$. Figure \ref{fig:filter_cusp}b shows a single pixel spectrum obtained from this technique with a $^{55}$Fe radioisotope. The FWHM at 5.9 keV Mn K$_\alpha$ line is measured to be around 140 eV. Note that, in this method, due to the increased weight of the samples close to the transition, which is essential for suppressing low frequency correlated noise, the effective integration time is reduced (bandwidth widens). Consequently, the thermal noise worsens, resulting in a moderate level of improvement in the overall noise yield.  

\subsubsection{Double-baseline filtering}
\begin{figure}
    \centering
      \includegraphics[width=1\linewidth]{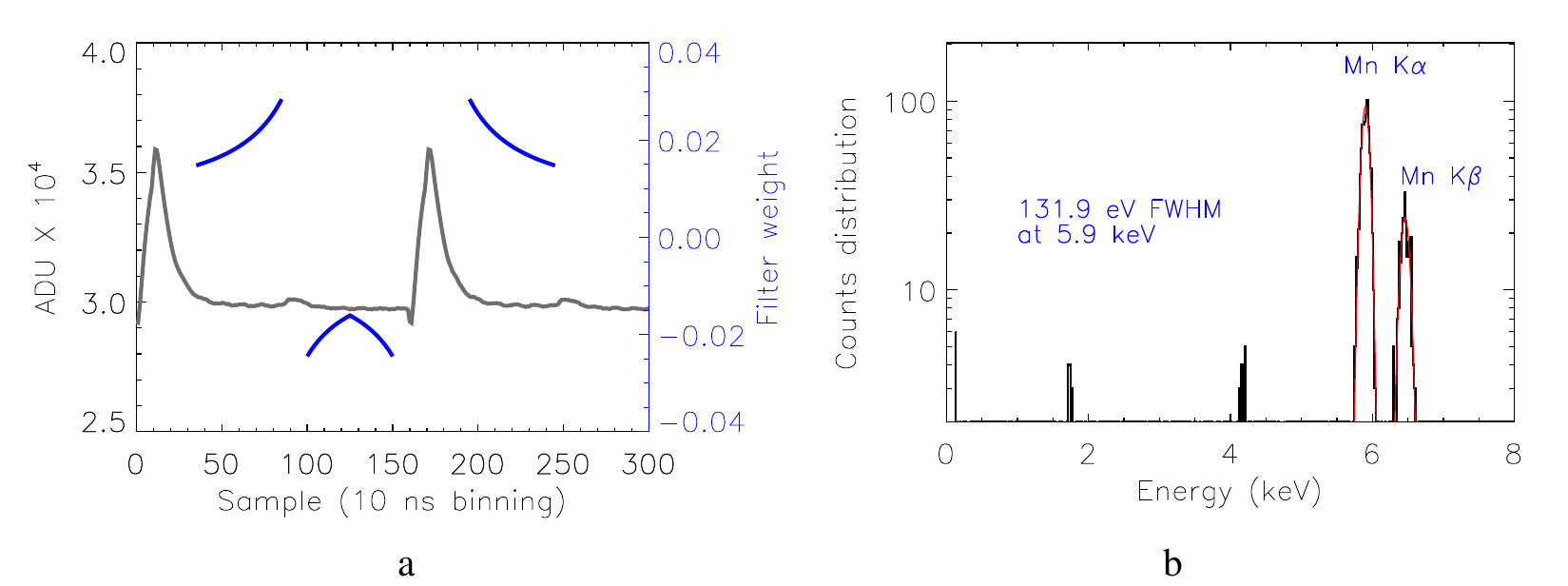}
    \caption{Results of double baseline digital filtering: (a) the blue lines (plotted against the right axis) show the weights of the double baseline cusp filter over the baseline and signal samples of the SiSeRO waveform (plotted against the left axis). With this filtering technique, the read noise is improved to 4.5 $e^{-}_{RMS}$. (b) An example spectrum obtained using the digital double baseline CDS filter. It shows the Mn K$_\alpha$ (5.9 keV) and K$_\beta$ (6.4 keV) lines from a $^{55}$Fe radioactive source for
single-pixel (grade 0) events. The FWHM is measured to be around 132 eV at 5.9 keV.}
    \label{spec_dbl_baseline}
\end{figure}

Ideally, in order to achieve minimum noise at the output, the sample size in the CDS should be maximized (minimum filter bandwidth) such that the thermal noise is at its minimum, but at the same time, the filter peak frequency should be high to ensure good 1/f noise suppression.  
To improve on the traditional filtering, we experimented by including baseline samples on either side of the pixel signal to calculate the pixel baseline level for CDS. This new baseline level not only uses a larger number of samples and therefore has lower white noise contribution, but it can be interpreted as the interpolation of the baseline value at the signal midpoint and therefore, the effect of slow moving variations are cancelled out, resulting in an improved overall noise performance. This approach works only because, SiSeROs do not suffer from reset or KTC noise and such, the baseline (i.e. the transistor state with an empty buried back gate) is expected to be always the same.  \cite{Hermannetal2022}.
Here the baseline and signal values can be estimated using either box-averaging or cusp-averaging of the samples. In case of cusp-averaging, we compute the signal values by employing double-sided cusp as shown in Fig. \ref{spec_dbl_baseline}a.
The read noise is measured to be around 4.5 $e^{-}_{RMS}$ which is around 25 \% improvement from the box CDS. Figure \ref{spec_dbl_baseline}b shows a single pixel spectrum obtained from this technique from a $^{55}$Fe radioactive source.      
Because of the low noise output, the energy resolution measurements also show a significant improvement, with FWHM at 5.9 keV Mn K$_\alpha$ line measured to be around 132 eV.

\subsubsection{Estimation and subtraction of correlated noise}

\begin{figure}
    \centering
         \includegraphics[width=1\linewidth]{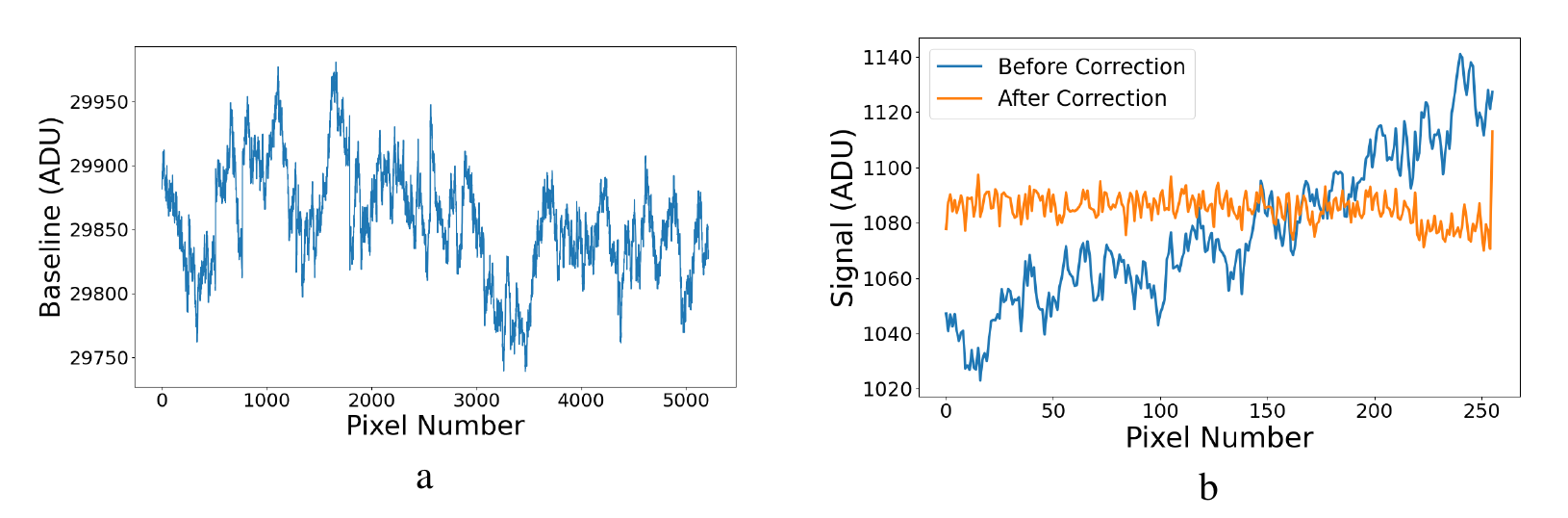}
    \caption{1/f noise filtering: (a) average baseline levels, calculated from the baseline samples of a pixel array (from a few consecutive array) are plotted against the respective pixels. Each pixel takes 1.6 $\mu$s to readout. From the baseline time series, we calculate the embedded noise frequencies using inverse fast Fourier transform (FFT) algorithm. The baseline values are recalculated from the FFT series at the signal times after shifting the series by appropriate amounts. (b) Measured signal charge (baseline - signal levels) for an array of pixels is shown before and after the 1/f baseline correction (in blue and orange respectively). The last 35 pixels at the end of each row represent the overscan pixels. The read noise is found to be around 4.7 $e^{-}_{RMS}$ with this method with an  FWHM of 132 eV at 5.9 keV Mn K$_\alpha$ line (not shown here).}
    \label{fig:1/f_method}
\end{figure}

In this method, we first calculate the baseline values for all the pixels in a row either by box-averaging or cusp-averaging the baseline samples. This is shown in Fig. \ref{fig:1/f_method}a where baselines computed for a few consecutive rows are shown together. Note that each pixel is 1.6 $\mu$s long. The effect of both the low- and the high-frequency noise contributions can be seen in the baseline time series. 

 While the baseline points are an approximation of the low frequency noise close to the signal points, the actual baseline will change slight in the time difference between the baseline and the signal. So subtracting the measured baseline will not do a perfect suppression of the low frequency noise. We could estimate the actual baseline at the time point of the signal by looking at the time series of successive baselines and estimate (interpolate) the baseline for the time points of the signal measurement. This operation is best performed in the frequency domain, and we use the following steps:
 \begin{enumerate} 
     \item Estimate the noise frequency distribution by applying a Fast Fourier Transform (FFT) on the measured time series of the baselines of a sequence of pixels (Fig. \ref{fig:1/f_method}a). 
     \item  Apply a phase shift to every frequency in the spectrum that is equivalent of the time difference between baseline and signal within a pixel. 
     \item  recalculate the baselines at the time point of the pixel signals by applying an inverse FFT to the phase shifted spectrum.
     \item The new baselines are then subtracted from the respective pixel signal levels.  
\end{enumerate}
In Fig. \ref{fig:1/f_method}b, the orange line represents the charge measured using this method while the blue line shows the same charge signals without this correction. The analysis is applied for each pixel row separately to obtain the final 1/f corrected image. In Fig. \ref{fig:box_cusp}, the noise yield measured using this method is shown in yellow circles along with the box and cusp filters (in blue and red respectively). Unlike the box or cusp filtering, the noise continues to decrease with the increase in sample size because of superior 1/f noise suppression. Using this technique the read noise is measured to be around 4.7 $e^{-}_{RMS}$. 
We found the FWHM at 5.9 keV Mn K$_\alpha$ line to be around 132 eV. Effectiveness of the method depends on the Nyquist frequency (in this case 312 KHz) and the 1/f corner frequency. A faster sampling frequency might also remove the fast moving noise signals, however, at the expense of higher thermal noise. This method involves various computational algorithms like the Fourier transforms, FFT shifts, filter windowing, optimization of which might improve the noise performance further and is currently under study.     


\section{Summary and future plans}\label{summary}
 
The SiSeRO amplifier, developed by MIT Lincoln Laboratory, is a novel technology for the output stage of X-ray CCDs that can in principle provide very low noise and high speed performance. 
In our previous paper \cite{chattopadhyay22_sisero}, we characterized the first generation SiSeRO devices obtaining a read noise of 15 $e^{-}_{RMS}$ and a FWHM energy resolution of 230 eV 
at 5.9 keV. The first prototypes used surface-channel transistors, where trapping and de-trapping of charge carriers from the silicon interface states degrades the overall noise performance. 
In this work, we have characterized a new SiSeRO device employing a buried transistor channel. We obtained significant improvements in the noise (around 6 $e^{-}_{RMS}$) and spectral 
performance (around 140 eV at 5.9 keV) at 625 kpixel/s readout speed. We also explored methods for 1/f correlated noise suppression using optimized filters for digital CDS instead of a 
default box filter. We achieved significant improvements in noise performance with the use of these tools: the spectral resolution obtained was 132 eV at 5.9 keV with read noise around 4.5 $e^{-}_{RMS}$. 

Future work will extend our development of SiSeRO devices to larger arrays with parallel readouts; the first examples will be 
fabricated shortly. We have already developed an ASIC-based readout system usable with such devices, the details of which 
can be found in Herrmann et al. 2020\cite{herrmann20_mcrc} and Orel et al. 2022 \cite{Oreletal2022}. Another important 
feature of SiSeRO devices is the capability to perform repetitive non-destructive readout (RNDR), a technique that holds 
the potential to reduce readout noise well below the 1/f barrier and deep into the sub-electron regime. A preliminary 
demonstration of the concept was presented in Chattopadhyay et al. 2022 \cite{Chattopadhyayetal2022}. A forthcoming paper 
will present our latest results.  


\subsection* {Acknowledgments}
This work has been supported by NASA grants APRA 80NSSC19K0499 ``Development
of Integrated Readout Electronics for Next Generation X-ray CCDs” and SAT
80NSSC20K0401 ``Toward Fast, Low-Noise, Radiation-Tolerant X-ray Imaging Arrays for
Lynx: Raising Technology Readiness Further.”




\end{spacing}
\end{document}